\pdfoutput=1
\documentclass[twocolumn,showpacs,floatfix,prl]{revtex4}

\usepackage[final]{graphicx}
\DeclareGraphicsExtensions{.pdf}
\usepackage{amsmath,amssymb,bm}
\usepackage{float}

\newcommand{\bk}{{\bf k}}
\newcommand{\bp}{{\bf p}}

\newcommand{\br}{{\bf r}}

\newcommand{\bB}{{\bf B}}
\newcommand{\bA}{{\bf A}}

\newcommand{\bE}{{\bf E}}
\newcommand{\bj}{{\bf j}}
\newcommand{\bn}{{\bf \hat{n}}}

\newcommand{\cH}{{\cal H}}
\newcommand{\cT}{{\cal T}}

\newcommand{\bsig}{{\bm \sigma}}
\newcommand{\bpi}{{\bm \pi}}

\begin{document}

\title{Wormhole Effect in a Strong Topological Insulator}

\author{G. Rosenberg, H.-M. Guo and M. Franz}
\affiliation{Department of Physics and Astronomy, University of
British Columbia,Vancouver, BC, Canada V6T 1Z1}

\begin{abstract}
An infinitely thin solenoid carrying magnetic flux $\Phi$ (a `Dirac
string') inserted into an ordinary band insulator has no significant
effect on the spectrum of electrons. In a strong topological
insulator, remarkably, such a solenoid carries protected gapless
one-dimensional fermionic modes when $\Phi=hc/2e$. These modes are
spin-filtered and represent a distinct bulk manifestation of the
topologically non-trivial insulator. We establish this `wormhole'
effect by both general qualitative considerations and by numerical
calculations within a minimal lattice model. We also discuss the
possibility of experimental observation of a closely related effect in
artificially engineered nanostructures.
\end{abstract}

\date{\today}

\maketitle

Surface electrons in a strong topological insulator (STI)
\cite{mele1,moore1,fu2,roy1,zhang1,cava1,cava2,ylchen1} form a gapless
helical liquid, protected by time reversal symmetry $(\cT)$ through
the topological invariants that characterize the bulk band
structure. When $\cT$ is broken, which may be accomplished by coating
the surface with a ferromagnetic film, the helical liquid transforms
into an exotic insulating state characterized by a precisely
quantized Hall conductivity
\begin{equation}\label{sig} \sigma_{xy}=\left(n+{1\over
2}\right){e^2\over h},
\end{equation}
with $n$ integer. This result follows from the microscopic theory of
the surface state \cite{fu2} and also from the effective
electromagnetic action describing the bulk of a topological insulator,
which contains the axion term \cite{qi1,essin1}. Although it might not
be possible to measure this `fractional' quantum Hall effect in a
transport experiment \cite{dhlee1}, Eq.\ (\ref{sig}) is predicted to
have observable physical consequences, such as the low-frequency
Faraday rotation \cite{qi1} and the image magnetic monopole effect
\cite{qi2}.

\begin{figure}[b]
\includegraphics[width=7.2cm]{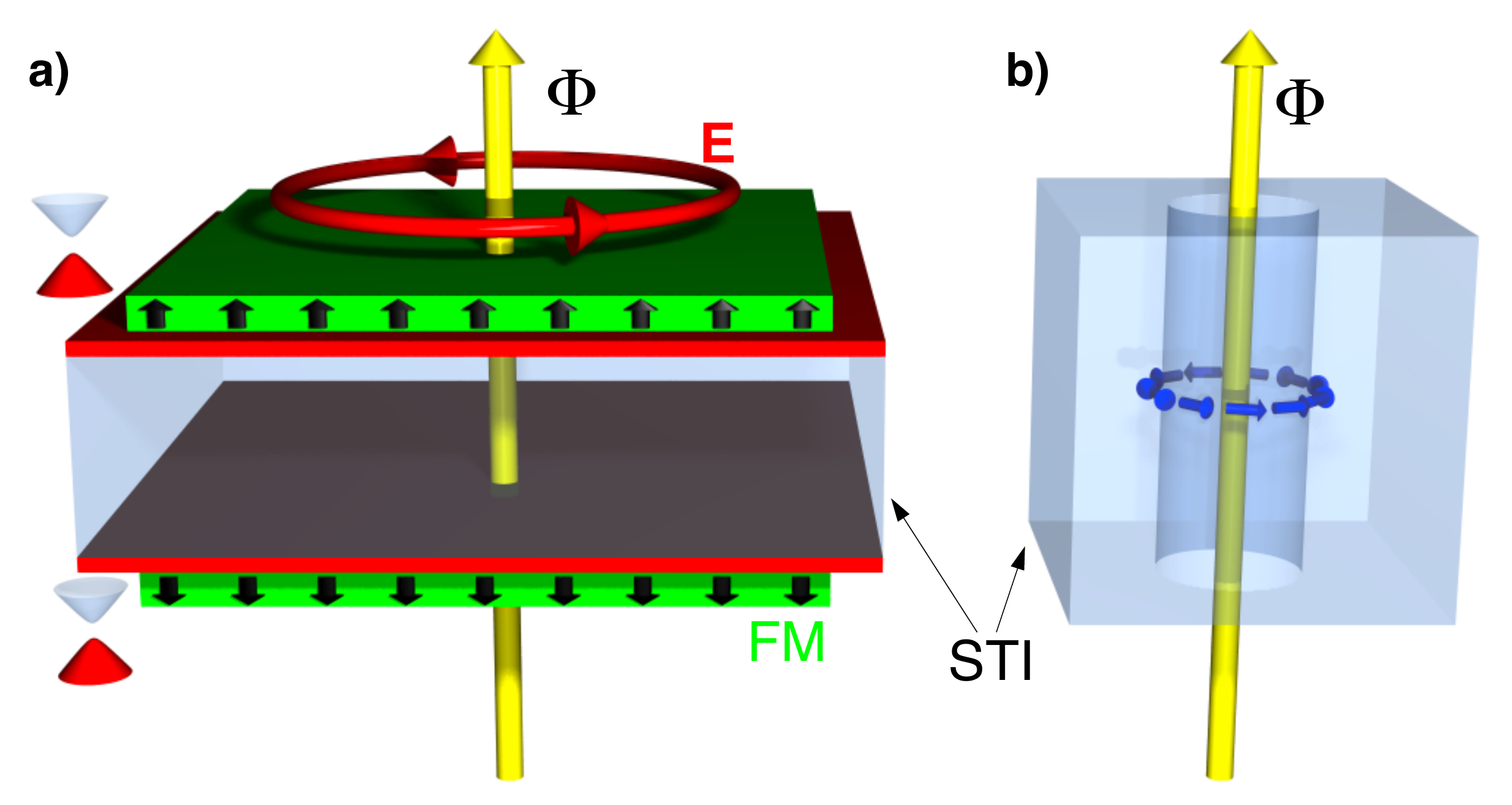}
\caption{(Color online) a) Topological insulator coated with a
ferromagnetic (FM) film. The flux tube employed in Laughlin's argument
and the induced electric field are indicated. b) Flux tube threading
a cylindrical hole in a STI. Arrows illustrate the helical
spin state for upward moving electrons (for down-movers the arrows
are reversed). } \label{fig1}
\end{figure}

It is instructive to apply Laughlin's flux insertion argument
\cite{laughlin1} to the STI surface described by Eq.\ (\ref{sig}). This
argument was devised to establish the fractional charge of
quasiparticles in fractional quantum Hall liquids (FQHL)
\cite{laughlin2} and involves the adiabatic insertion of an infinitely
thin solenoid carrying magnetic flux $\Phi(t)$ into the system, as
illustrated in Fig.\ \ref{fig1}a. As the flux is ramped up from 0 to
$\Phi_0=hc/e$, a circumferential electric field is generated in accord
with Faraday's law $\nabla\times\bE=-(1/c)(\partial\bB/\partial
t)$. This induces a Hall current on the STI surface 
$\bj=\sigma_{xy}(\bE\times\hat{z})$ which brings electric charge
\begin{equation}\label{dQ} \delta Q=\sigma_{xy}{\Phi_0\over
c}=\left(n+{1\over 2}\right)e
\end{equation}
to the solenoid.  Since the flux tube carrying a full flux quantum
$\Phi_0$ can be removed from the electronic Hamiltonian by a gauge
transformation, one concludes, as in FQHL, that an excitation with
fractional charge (\ref{dQ}) must exist.  This finding stands in
contradiction to the well established microscopic theory of these
surface states given by an odd number of massive Dirac Hamiltonians
\cite{fu2}. Elementary excitations of a massive Dirac Hamiltonian
are particle-hole pairs which are charge neutral. Yet, this same Dirac
Hamiltonian exhibits Hall conductivity (\ref{sig}), which, through
Laughlin's argument outlined above, implies fractionally charged
quasiparticles.

The resolution to this paradox comes from the realization that the
quantum Hall state realized on the surface of a STI is inextricably
linked to the bulk of the STI. Laughlin's argument fails because the
flux tube inserted into the {\em bulk} of the STI is not inert. We
demonstrate below that when $\Phi=(s+1/2)\Phi_0$, with $s$ integer, the
flux tube carries topologically protected gapless fermionic modes and
forms a conducting quantum wire -- a `wormhole' -- along which the
accumulated surface charge can escape to another surface of the
sample. In the end, no net fractional charge is accumulated at the surface
and Laughlin's argument instead predicts, indirectly, a new effect
associated with a Dirac string in the bulk of a STI that we propose to
call a `wormhole effect'.

In the rest of this Letter we establish the wormhole effect, first by
an analytical calculation using the universal properties of the
surface states, and then by numerical calculations within a lattice
model of a STI. We discuss its physical properties, significance and
the possibility of experimental observation.

We begin by considering a bulk STI with a cylindrical hole of radius
$R$ threaded by magnetic flux $\Phi=\eta\Phi_0$ with $0\leq\eta<1$ as
illustrated in Fig.\ \ref{fig1}b. By solving the Dirac equation for
the surface electrons we show that a gapless state exists when
$\eta={1\over2}$ and persists in the limit $R\to 0$. According to
Ref.\ \cite{mirlin1} electron states on a curved surface of a STI,
characterized by a normal unit vector $\bn$, are described by a Dirac
Hamiltonian of the form
\begin{equation}\label{dir1} \cH={1\over 2}v\bigr[\hbar\nabla\cdot\bn
+ \bn\cdot(\bp\times\bsig) + (\bp\times\bsig)\cdot\bn\bigl]
\end{equation}
where $v$ is the Dirac velocity, $\bp=-i\hbar\nabla$ is the momentum
operator and $\bsig=(\sigma_1,\sigma_2,\sigma_3)$ is the vector of
Pauli spin matrices. The magnetic flux is included by
replacing $\bp$ with $\bpi=\bp-(e/c)\bA$, where
$\bA=\eta\Phi_0(\hat{z}\times\br)/2\pi r^2$ is the vector
potential. For a cylindrical inner surface
$\bn=-(\cos{\varphi},\sin{\varphi},0)$, the Hamiltonian
(\ref{dir1}) becomes, in cylindrical coordinates and taking
$v=\hbar=1$,
\begin{equation}\label{dir2} \cH_k=-{1\over 2R}+
\begin{pmatrix} {1 \over R}(i\partial_\varphi+\eta) & -ike^{-i\varphi}
\\ ike^{i\varphi} & -{1 \over R}(i\partial_\varphi+\eta)
\end{pmatrix}.
\end{equation}
We assumed a plane-wave solution $e^{ikz}$ along the cylinder axis and
replaced $-i\partial_z\to k$.

The eigenstates of $\cH_k$ are of the form
\begin{equation}\label{psi1} \Psi_k(\varphi)=\begin{pmatrix} f_k \\
e^{i\varphi}g_k \end{pmatrix} e^{i\varphi l}
\end{equation}
with $l$ integer. The spinor $\tilde\Psi_k=(f_k,g_k)^T$ is an eigenstate
of $\tilde\cH_{kl}=\sigma_2k-\sigma_3(l+{1\over2}-\eta)/R$ with an
energy eigenvalue
\begin{equation}\label{ekl} E_{kl}=\pm v\hbar\sqrt{k^2+{(l+{
1\over2}-\eta)^2\over R^2}}.
\end{equation}
For a generic strength of the magnetic flux the spectrum of electrons
along the cylindrical surface shows a gap
\begin{equation}\label{gap} \Delta={2v\hbar\over
R}\left|{1\over2}-\eta\right|.
\end{equation}
When $\eta={1\over 2}$, {\em i.e.}~at half flux quantum, the $l=0$
mode becomes gapless, $E^{\eta=1/2}_{k0}=\pm v\hbar |k|$, independent
of the hole radius $R$. This is the wormhole effect introduced above.
Physically, the necessity of the flux for the gapless state to occur 
stems from the Berry's phase $\pi$ acquired by electron spins in the helical
state  depicted in Fig.\ \ref{fig1}b. The gapless state occurs at half flux
quantum when the Aharonov-Bohm phase exactly cancels the spin Berry's phase.

We observe that the system remains $\cT$-invariant in the presence of
a half flux quantum threading the hole. Therefore, the gapless state
is topologically protected against any weak perturbation that respects
$\cT$ and does not close the bulk gap. Specifically, it should be
robust against weak non-magnetic disorder as well as any smooth
deformation of the hole. Our numerical simulations, presented below,
provide support for this topological protection.
\begin{figure}[t]
\includegraphics[width=8.2cm]{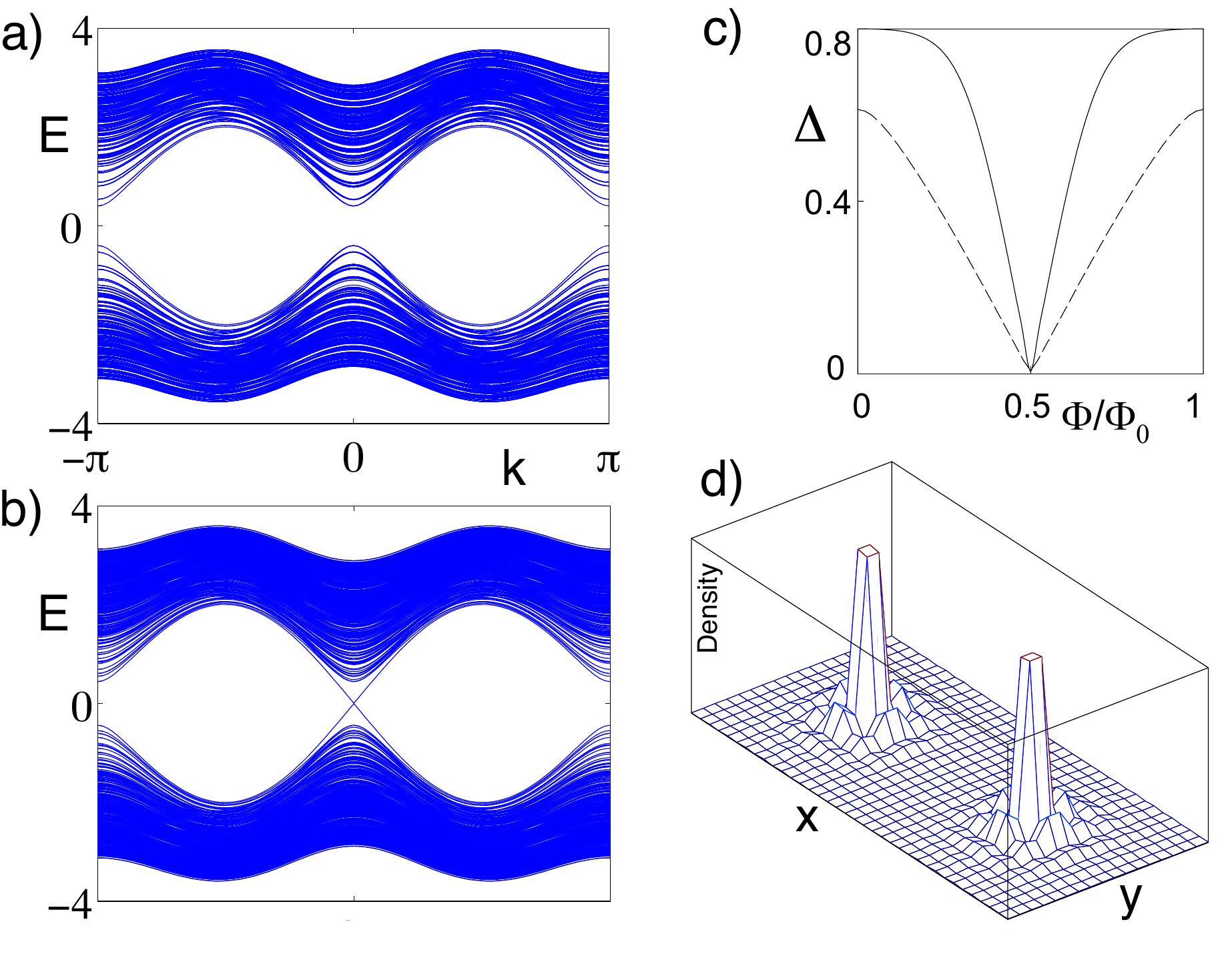}
\caption{(Color online) Band structure of $36\times 18$ STI sample
(periodic boundary conditions), infinite along the $z$-direction with
two flux tubes carrying flux $\eta\Phi_0$ and $-\eta\Phi_0$ inserted,
a) $\eta=0$, b) $\eta=1/2$. Panel c) shows dependence of the spectral
gap on the flux for thin flux tubes threading a lattice plaquette
(solid line), and thicker flux tube threading a $3\times 3$
rectangular hole (dashed line). d) Wavefunction amplitude for a
low-lying state extended along the flux tubes. Parameters used here
are $t=0.2$, $\epsilon=0.8$.  } \label{fig2}
\end{figure}

We now study the wormhole effect using a concrete lattice model of a
topological insulator which we solve by exact numerical
diagonalization. In order to keep the computational difficulties at
a minimum we consider a simple model on a cubic lattice discussed
previously \cite{qi1,rosenberg1}. This minimal model has two electron
orbitals per lattice site, denoted $c$ and $d$, and is defined by the
momentum space Hamiltonian $H=\sum_\bk\Psi_\bk^\dagger{\cal
H}_\bk\Psi_\bk$ with
$\Psi_\bk=(c_{\bk\uparrow},c_{\bk\downarrow},d_{\bk\uparrow},d_{\bk\downarrow})^T$,
\begin{equation}\label{hk} {\cal
H}_\bk=-2\lambda\sum_\mu\tau_z\sigma_\mu\sin{k_\mu}+\tau_xm_\bk,
\end{equation}
and $m_\bk=\epsilon-2t\sum_\mu\cos{k_\mu}$. Here $\tau_\mu$ and
$\sigma_\mu$ are Pauli matrices in orbital and spin space,
respectively, with $\mu=x,y,z$. The system defined by $H$ is invariant
under time-reversal and spatial inversion. The spectrum of excitations
has two doubly degenerate bands,
\begin{equation}\label{ek}
E_\bk=\pm\sqrt{4\lambda^2(\sin^2{k_x}+\sin^2{k_y}+\sin^2{k_z})+m_\bk^2}.
\end{equation}
At half filling, depending on the values of the parameters $\lambda$, $t$,
$\epsilon$, the system can be a trivial insulator, as well as a strong
and weak topological insulator (WTI) \cite{qi1,rosenberg1}. Below,
unless stated otherwise, we work with parameters $2t<\epsilon<6t$,
corresponding to a STI phase characterized by the $Z_2$ invariant
$(1;000)$. All energies are expressed in units of $\lambda$ which we
take equal to $1$.

To look for gapless propagating modes along a flux tube we first
consider a sample infinite in the $z$-direction with a rectangular
base containing $2L\times L$ sites. Two straight flux tubes carrying
fluxes $\eta\Phi_0$ and $-\eta\Phi_0$ along the $z$-direction are
positioned a distance $L$ apart on the $x$ axis. Since the total flux
threading the system is zero for this arrangement we may use periodic
boundary conditions along $x$ and $y$ and thus eliminate gapless modes
that would otherwise reside on surfaces. Results for $L=18$ are
displayed in Fig.\ \ref{fig2}. Without the flux $(\eta=0)$ the system
shows a spectral gap. For $\eta>0$ subgap states appear near
$k=0$. When $\eta={1\over2}$ a gapless mode exists along each of the
flux tubes.

In addition to thin flux tubes that thread an elementary plaquette we 
also considered flux threading a larger rectangular hole
in the sample. Results for this case are similar; a gapless mode
appears when $\eta={1\over2}$ and the dependence of the gap on $\eta$
(Fig.\ \ref{fig2}c) now more closely resembles that given in Eq.\
(\ref{gap}). In general gapless modes persist for
any size and shape of the hole, as long as it is threaded by a half flux quantum.

We performed similar calculations for other topological phases
occurring in the same model. In a $(1;111)$ STI that occurs when
$-6t<\epsilon<-2t$ a single gapless mode (per flux tube) exists, now
located near $k=\pi$. In WTI phases an even number of gapless modes per
flux tube appear. For a straight flux tube along direction $\bn$ we
find two gapless modes (one at $k=0$ and one at $k=\pi$) when
$\bn\cdot{\bm\nu}\neq 0$ and zero otherwise. Here
${\bm\nu}=(\nu_1\nu_2\nu_3)$. Finally, we have verified that in a
trivial $(0;000)$ insulator, that occurs for $|\epsilon|>6t$, no gapless
modes appear for any direction or strength of flux tube.

\begin{figure}[t]
\includegraphics[width=8.2cm]{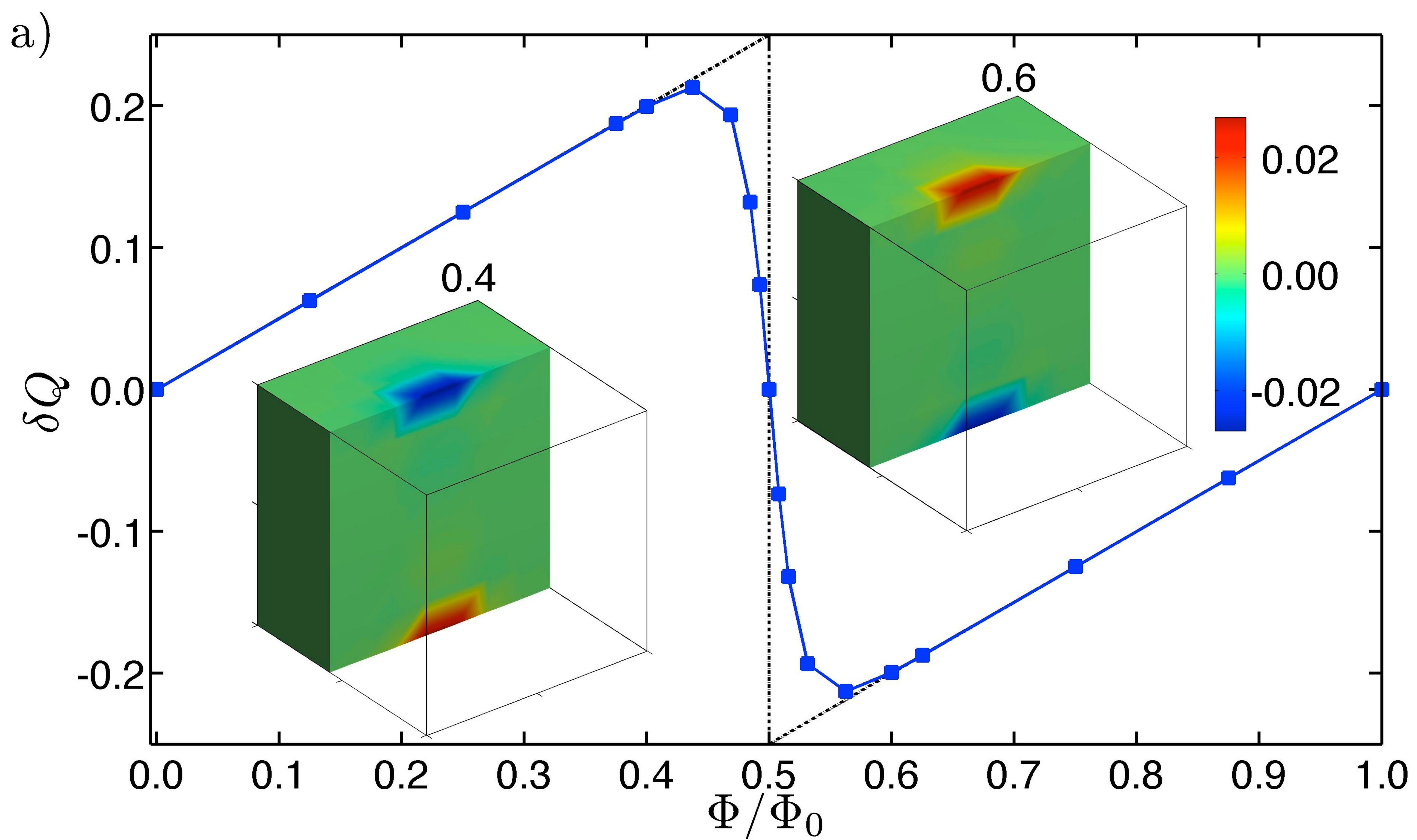}
\includegraphics[width=8.2cm]{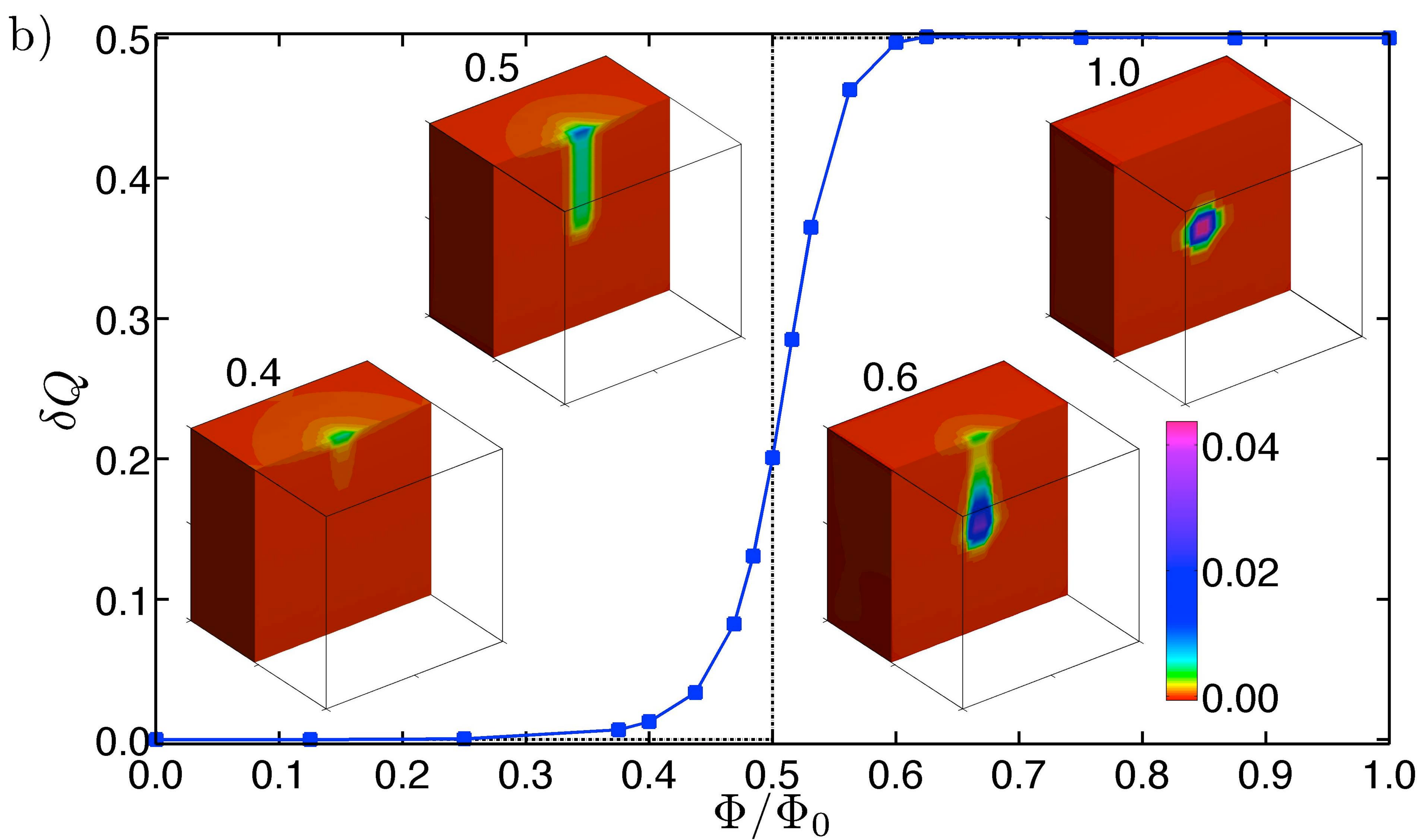}
\caption{(Color online) a) Charge $\delta Q$ (in units of $e$) induced
in the lower half of the sample by a flux tube carrying flux
$\Phi$. Insets show the charge density at indicated values of flux. b)
Charge $\delta Q$ in the sphere of radius $r_0$ centered at the
monopole as a function of monopole strength. We have chosen
$r_0=9.097$ so that the sphere is large enough to enclose the expected
charge $e/2$ for the unit monopole projecting flux $\Phi_0$. In both
panels a cube-shaped sample with $20^3$ lattice sites and parameters
$t=1$, $\epsilon=4$ is used, $\Omega_S=1,0$ in a) and b) respectively. $\delta Q$ is defined relative to
the $\eta=0$ situation.  } \label{fig3}
\end{figure}
Using our model with a single flux tube in a geometry with open
boundary conditions it is possible to visualize the flow of charge at
the intermediate steps of Laughlin's flux insertion argument. To
this end we consider a cube of size $L^3$ and supplement the
Hamiltonian (\ref{hk}) with a surface magnetization term
\begin{equation}\label{hs} H_S=-\Omega_S\sum_{j\in{\rm
surf}}\hat{\br}_j\cdot\left(\Psi^\dagger_j\bsig\Psi_j\right).
\end{equation}
Here $\hat{\br}_j$ represents the unit vector pointing outward from
the origin located at the cube's center and $\Omega_S$ is the surface
magnetization strength. $H_S$ breaks $\cT$ at the sample surface and
a gap of size $\sim 2|\Omega_S|$ opens up in the spectrum of the
surface states. Figure \ref{fig3}a shows the evolution of charge
$\delta Q$ accumulated near the intersection of the flux tube with the
magnetized surface as a function of
$\eta=\Phi/\Phi_0$. For small $\eta$ we observe $\delta Q={1\over
2}e\eta$, consistent with the fractional Hall conductivity
$\sigma_{xy}=e^2/2h$ expected on the basis of Eq.\ (\ref{sig}). At
$\eta={1\over 2}$ a charge $e/2$ travels along the flux tube and
combines with the negative charge that has built up on the opposite
surface. For $\eta>{1\over 2}$ the charge $\delta Q$ grows again at
the rate controlled by $\sigma_{xy}$ until it reaches $\delta Q=0$ at
$\eta=1$. As already mentioned above, a Dirac string carrying a full
flux quantum $\Phi_0$ can be removed by a gauge transformation and the
above evolution is thus consistent with the expectation that this
weakly interacting system returns to the original configuration at
the end of a full cycle.

Figure \ref{fig3}b displays a modified arrangement with the flux tube terminated by a
magnetic monopole located at the center of the sample. This furnishes
a realization of the Witten effect \cite{witten1} in a STI
\cite{rosenberg1}. As a function of increasing $\eta$, the charge first
accumulates at the intersection of the flux tube and the surface. At
$\eta={1\over 2}$ a charge $e/2$ travels along the wormhole to the
monopole, the corresponding charge density clearly visible in the
inset to Fig.~\ref{fig3}b. At $\eta=1$ the flux tube becomes invisible
but the $e/2$ charge remains bound to the monopole as expected on the
basis of general arguments \cite{witten1,rosenberg1}. 

We performed similar calculations for flux tubes of various shapes and in 
the presence of weak non-magnetic disorder. In all cases we found low-energy
modes associated with the $\eta={1\over2}$ flux tube confirming the topological robustness of the wormhole effect.

We now address the possibility of experimental detection of the
wormhole effect predicted in this Letter. In a real physical system it
is not possible to confine magnetic flux to an area of size
comparable to the crystal lattice spacing as would be necessary to
probe the wormhole effect in its pure form. However, it should be
possible to observe a closely related effect in a nanoscale hole
fabricated in a STI crystal with a uniform magnetic field applied
parallel to its axis, Fig.\ \ref{fig1}b. Sweeping the magnetic field
strength will result in a periodic variation of the conductance along
the hole with minima at $n\Phi_0$ and maxima at $(n+1/2)\Phi_0$ as the
excitation spectrum oscillates between insulating and metallic. Such
variations should be observable experimentally if certain conditions
are met. First, the hole radius $R$ must be sufficiently large so that
several oscillations can be observed in the available range of the 
laboratory field $B$. This gives $R\gtrsim (N\Phi_0/\pi B)^{1/2}$ for
$N$ oscillations. Second, $R$ must be sufficiently small so that the
maximum spectral gap Eq.\ (\ref{gap}) is large compared to $k_BT$, or
otherwise the oscillations in the conductance will be washed out by thermal
broadening. This gives $R\lesssim \hbar v/\sqrt{2}k_BT$. Taking
typical values $B=10$T, $N=10$, $v=5\times 10^5$m/s and $T=1$K yields
$36\ {\rm nm}\lesssim R \lesssim 450$ nm. Thus, the experimental challenge 
would lie in fabricating a sub-micron size hole (or an array of holes)
in a STI crystal or a thick film and measuring the conductance along the holes.

Oscillations with period $\Phi_0$ have been observed in recent
conductance measurements on Bi$_2$Se$_3$ single-crystal nanoribbons
(cross sections $6-10\times 10^{-15}$m$^2$, consistent with the above
bounds on $R$) in longitudinal magnetic field \cite{peng1}. In these,
the same effect as discussed above should occur for the topologically
protected states on the outer surfaces of the nanoribbon. However,
the observed positions of minima and maxima were {\em opposite} to
those predicted by our theory, suggesting that conductance in these
experiments is dominated by some competing effect. The oscillations
reported in Ref.\ 
\cite{peng1} clearly deserve a detailed theoretical study.

The wormhole effect introduced here is fundamentally different from the
gapless modes predicted to exist along the core of a
crystal dislocation in topological insulators \cite{ran1}. These latter
modes depend solely on the {\em weak} invariants $(\nu_1\nu_2\nu_3)$
whereas the wormhole effect depends on the more robust {\em strong}
invariant $\nu_0$. In this sense the wormhole effect is inherently
three-dimensional while the gapless modes associated with a
dislocation are more closely related to the zero-modes in
2-dimensional topological (spin-Hall) insulators with solitonic
defects \cite{ran2,qi7}.  Mathematically, the wormhole effect is
related to the protected one-dimensional modes predicted to exist
along vortex lines in the order parameter characterizing a topological
Mott insulator \cite{ran3}. Unfortunately, no topological Mott
insulators are known to exist at present, although according to Ref.\
\onlinecite{pesin1} the pyrochlore compounds $A_2$Ir$_2$O$_7$ ($A=\mbox{Pr,Eu}$)
may exhibit this behavior. 

The wormhole effect studied in this Letter represents a distinct {\em bulk}
manifestation of the unusual electron properties in a strong
topological insulator. Its existence resolves a conceptual dichotomy
that arises when Laughlin's argument is applied to the magnetized STI surface and exemplifies a unique bulk-surface correspondence inherent to STIs.
A closely related counterpart of the wormhole effect should be observable in artificially engineered nanostructures fabricated from available STIs.

\emph{Acknowledgment} --- The authors are indebted to J. Moore, I. Garate,
and C. Weeks for stimulating discussion, and especially to B. Seradjeh
for help in formulating the experimental proposal. Support for this
work came from NSERC, CIfAR and The China Scholarship Council.

\emph{Note added} --- When this work was close to completion we became
aware of a preprint \cite{mirlin1} which has identified, in a
different context, the gapless modes on a surface of a STI cylinder
threaded by flux $\Phi_0/2$.


\end{document}